\documentclass[a4paper]{article}
\usepackage{amsfonts}
\usepackage{amsmath,amssymb,amsthm}
\usepackage{latexsym}
\usepackage[english]{babel}
\usepackage{graphicx}

\begin{document}
\title{On peakons and Q-balls in the baby Skyrme model}
\author{Jakub Lis \thanks{lis@th.if.uj.edu.pl}\\
Department of Field Theory,\\Institute of Physics at the Jagiellonian University\\
\small{Reymonta 4, 30-059 Cracow, Poland}
}
\maketitle

\begin{abstract}
In this paper we investigate the Q-ball Ansatz in the baby Skyrme model. First,  the appearance of peakons, i.e. solutions with  extremely large absolute values of the second derivative at  maxima, is  analyzed. It is argued that such solutions are intrinsic to the baby Skyrme model and do not depend on the detailed form of a potential used in calculations. Next, we concentrate on compact non spinning Q-balls. We show the failure of a small parameter expansion in this case. Finally, we explore the existence and parameter dependence of Q-ball solutions.
\end{abstract}

\section{Introduction}
The baby Skyrme model is an important member of the family of nonlinear field theories. Although it was proposed  as a simplified version of the Skyrme model, it is used to model some real-world systems, see~\cite{adam},~\cite{speight1} and references therein. From the formal point of view, the theory describes dynamical maps between two spheres $S_{2}$. This gives rise to a non-trivial topological structure and most efforts were done in this domain. However, in this paper we shall put the topological questions aside since the trivial topological sector is also  quite interesting, see~\cite{zakrzewski}.  If a customary field potential is chosen, there is an additional, apart from the topological charge,  quantity untouched by the time evolution. In the scalar field formulation it is simply the $U(1)$ charge. Its presence validates the question about existence of Q-balls, solutions known in many theories to minimize the energy for a given value of charge.\\
The present  work aims at clarifying some findings described in~\cite{adam}.  In that paper C.~Adam~\emph{et al.} report on existence of both spinning and non-spinning  compact Q-balls in the baby Skyrme model with a V-shaped potential. Some of these solutions, called peakons,  have a peculiar shape -- the value of the second derivative at their maxima becomes extremely large. 
These observations come from numerics and some of them  lack satisfactory analytical explanation. In the present paper we shed some light on them. First, we identify a structure in equation for the Q-ball profile function responsible for the appearance of peakons.  Second, we describe qualitatively the existence domain of the non-spinning Q-balls in the  theory. On the way,  difficulties with the customary small parameter expansion in theories with non smooth field force are discussed.\\
We consider a sharp potential associated with the signum-Gordon model (see e.g.~\cite{ar}) giving rise to compact solutions and requiring a  matching procedure. Some objections to the potential has been risen in~\cite{speight2}. On the other hand,  the article~\cite{rosenau} may be read as an apologia for such potentials. The solutions described in the present paper are legitimate weak solutions to a  differential equation, a  discussion of this fact may be found in~\cite{ar3}. Another attitude toward the potential is also  possible. Then it holds as an idealization, where the close-to-the-vacuum effects are neglected. The  results in~\cite{lis} suggest that a regularization of the potential would not change seriously the below described solutions.\\
The paper is organized as follows.  After a short introductory section  we discuss the well-posedness problem of the equation for the Q-ball profile function. This issue is intimately connected to the appearance  of peakons.  In the next section we give an account of our futile attempt to formulate a small parameter expansion for Q-balls in the theory at hand. A natural and well working approximation is introduced in the section 5. It makes it possible to characterize the Q-ball dependence on parameters present in the equation. The last section  contains our conclusions and remarks.

\section{General settings}
Let us consider the following Lagrange density function:
\begin{equation}\label{lagrange}
\mathcal{L}=4 \frac{\partial_{\mu}\Phi\partial^{\mu}\bar{\Phi}}{\left(1+|\Phi|^{2}\right)^{2}}-8\beta \frac{\left(\partial_{\mu}\Phi\partial^{\mu}\bar{\Phi}\right)^{2}-\partial_{\mu}\Phi\partial^{\mu}\Phi\partial_\nu\bar{\Phi}\partial^{\nu}\bar{\Phi}}{\left(1+|\Phi|^{2}\right)^{4}}-\lambda \frac{|\Phi|}{\sqrt{1+|\Phi|^{2}}},
\end{equation}
where $\Phi$ is a complex scalar field, $\bar{\Phi}$ its complex conjugation and $|\Phi|^{2}=\Phi\bar{\Phi}$; $\beta>0$ and $\lambda>0$. The above form of the Lagrangian in three $(2+1)$ dimensional Minkowski space-time defines the baby Skyrme model in the scalar field formulation, for details see \cite{adam}. The last term in the Lagrangian is a field potential. In this paper we use the V-shaped one following \cite{adam}. The charge $Q$ coming from the $U(1)$ phase invariance is given by the formula
\begin{equation}\label{chargeA}
Q=Im\int d^{2}x \ \left[ 4 \frac{\Phi \partial_{0}\bar{\Phi}}{\left(1+|\Phi|^{2}\right)^{2}}+ \frac{16\beta\Phi}{\left(1+|\Phi|^{2}\right)^{4}} \left(\partial_{\nu}\bar{\Phi}\partial^{\nu}\bar{\Phi}\partial_{0}\Phi-\partial_{\nu}\Phi\partial^{\nu}\bar{\Phi}\partial_{0}\bar{\Phi}\right)\right].
\end{equation}
Let us stress  that this quantity is not related to the topological structure of the theory.\\
We shall consider  solutions of the model set by~(\ref{lagrange}) found with help of  the spinning Q-ball Ansatz:
\begin{displaymath}
\Phi(x)=e^{i\left(\omega t+n\phi\right)} f(r),
\end{displaymath}
where $\omega>0$, $(r,\phi)$ are  radial and angle coordinates, $f$ is a real valued function and finally $n$ is an integer. In~\cite{adam} it is pointed out, that for small amplitudes the theory reduces to the signum-Gordon model. This fact suggests writing the equation in a form corresponding to the \emph{on-shell} scaling symmetry present in the signum-Gordon theory. Rescaled variable $y=\omega r$ and function $\lambda g(y)=8 \omega^{2} f(r)$  should be appropriate and lead to some simplifications. In this setting we obtain the equation
\begin{eqnarray}\label{gen-eq}
\frac{1}{y}\frac{d}{d y}
\left[y g' \left(1+\frac{\kappa\gamma g^{2} \left(\frac{n^{2}}{y^{2}}-1\right)}
 {\left(1+\kappa^{2}g^{2}\right)^{2}}\right)\right]=
g\left(\frac{n^{2}}{y^{2}}-1\right)
\left(1+\frac{\kappa\gamma {g'}^{2}} {\left(1+\kappa^{2}g^{2}\right)^{2}}\right) \nonumber \\
+\frac{2 \kappa^{2} g}{1+\kappa^{2} g^{2}}
\left[{g'}^{2}-g^{2}\left(\frac{n^{2}}{y^{2}}-1\right) \right]
+sign(g)\sqrt{1+\kappa^{2}g^{2}},
\end{eqnarray}
where $\kappa=\lambda/8\omega^{2}$ and $\gamma=\lambda \beta$.  Thus, the equation depends on two combinations $\{\gamma, \ \kappa\}$ of the original three parameters $\{\lambda, \  \beta, \ \omega \}$. The function $sign(g)=1$ if $g>0$ and $sign(g)=0$ if $g=0$. We shall  be interested only in  nontrivial solutions with  $g>0$, hence we  omit the function $sign(\cdot)$ in what follows.  The solutions of physical importance satisfy the boundary conditions ensuring  finiteness of the energy and the charge. So, before we proceed with the analysis of the equation~(\ref{gen-eq}), let us write down the expressions for the energy and the $U(1)$ charge in terms of the rescaled variable~$y$ and function~$g$
\begin{eqnarray}\label{energy}
E=2\pi \kappa^{2}\int_{0}^{\infty} dy \ y \left\{
\frac{4}{(1+\kappa^{2}g^{2})^{2}}\left[{g'}^{2}+g^{2}\left(\frac{n^{2}}{y^{2}}+1\right)\right]+ \right.\\ 
\left. +\frac{4\gamma\kappa {g'}^{2}g^{2}}{\left(1+\kappa^{2}g^{2}\right)^{4}}\left(\frac{n^{2}}{y^{2}}+1\right)+
 \frac{8|g|}{\sqrt{1+\kappa^{2}g^{2}}} \right\} \nonumber 
\end{eqnarray}
and
\begin{equation}\label{charge}
Q=4 \pi \kappa^{5/2}\sqrt{\frac{8}{\lambda}}\int_{0}^{\infty} dy \ y \left[\frac{2 g^{2}}{(1+\kappa^{2}g^{2})^{2}}-\frac{\gamma \kappa g^{2}{g'}^{2}}{\left(1+\kappa^{2}g^{2}\right)^{4}}\right].
\end{equation}
In what follows we  vary $\kappa$ while $\gamma$ is kept constant. This choice of the control parameter is rather natural. For a given model the parameters $\lambda$ and $\beta$ (thus~$\gamma$) are set. The parameter $\kappa$ depends also on $\omega$ that characterizes initial data. Hence, it may be any real number.

\section{Appearance of peakons}
\begin{figure}
  \begin{center}
    \includegraphics[width=0.8\textwidth]{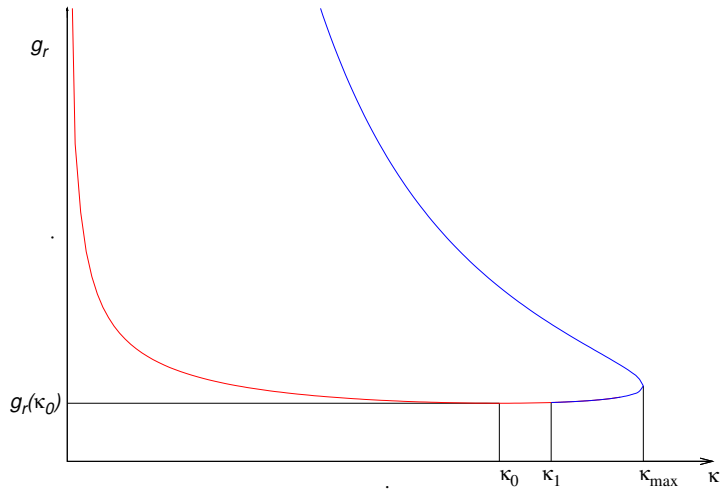}
    \caption{The structure of the roots of the relation~(\ref{gam0}) for given $\gamma$. The maximal value of $\kappa$ for which a real solution exists is  $\kappa_{max}=\gamma/4$. A minimal value $g_{r}$ for a given $\gamma$ is equal to $16/(\sqrt{27}\gamma)$ and is reached for $\kappa_{0}=3 \gamma/16$. For the value $\kappa_{1}$ it holds $\alpha_{2}=0$; the blue line (on the right from $\kappa_{1}$) marks $\alpha_{2}>0$ and the for red line (on the left from $\kappa_{1}$) $\alpha_{2}<0$. For small $\gamma$ the two colors meet on the upper branch, for large values of $\gamma$ on the lower one.} 
    \label{roots1}
  \end{center}
 \end{figure}

\begin{figure}
  \begin{center}
   \begin{tabular}{cc}
    \includegraphics[width=0.49\textwidth]{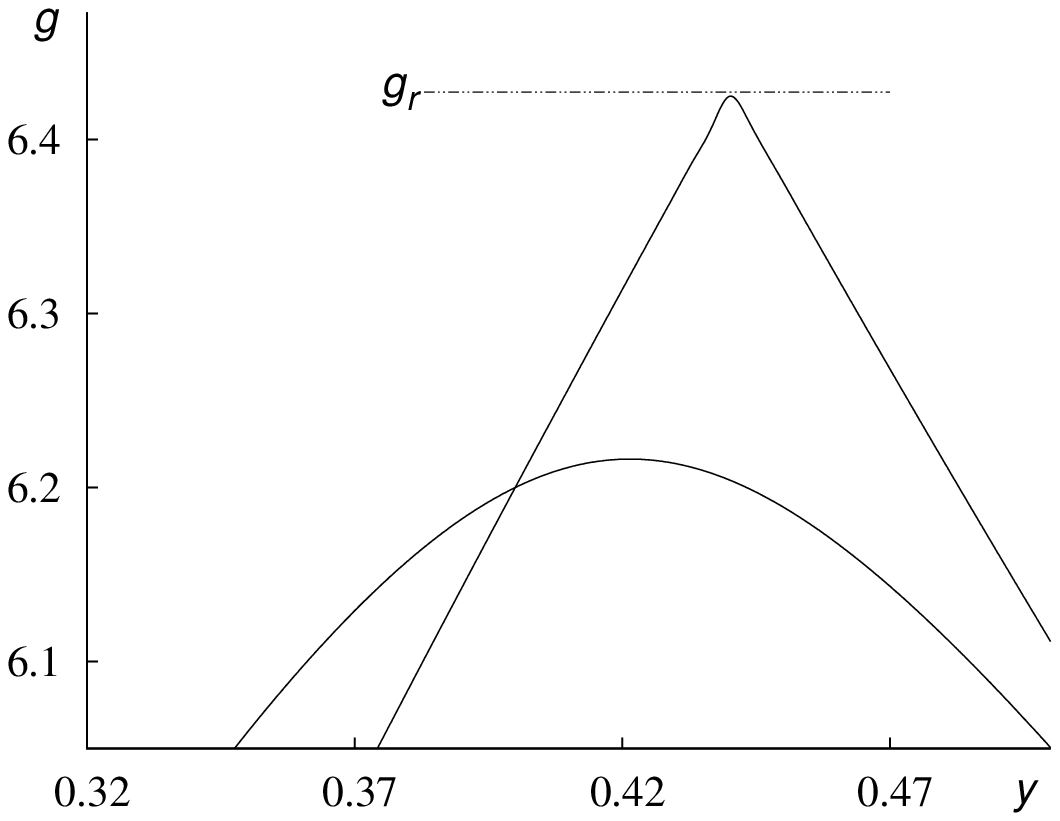} & \includegraphics[width=0.49\textwidth]{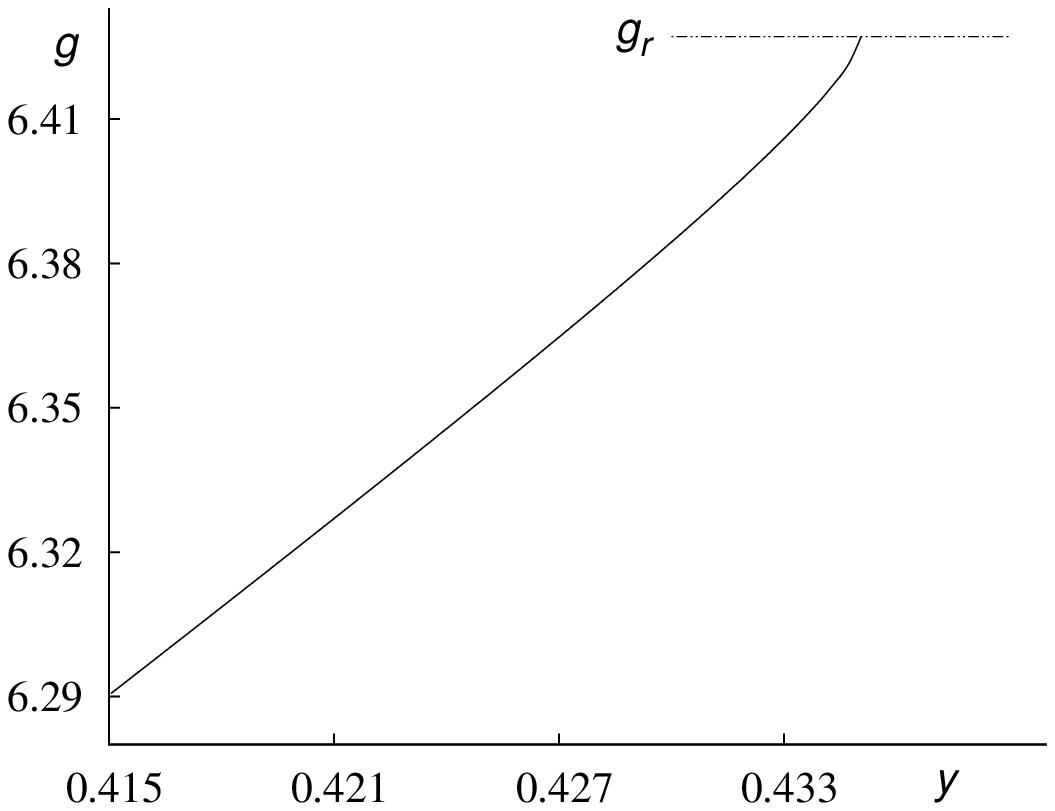}\\
    \end{tabular}
    \caption{Numerical solutions of the equation~(\ref{nto0}) with $\gamma=0.5$ and $\kappa=0.07$. This implies $g_{r}=6.427$ and $\alpha_{2}=15.34$. Initial data are set for $y=0.4$ and  $g =6.2$. On the right panel the derivative $g'$ of the solution approaching closer to $g_{r}$ is  $5.745$, the other solution starts with $g'=1.5$. The solution on the right panel is solved with $g'=6$.}
   \label{ilustr}
  \end{center}
 \end{figure}

Henceforth we consider the equation (\ref{gen-eq}) for $n=0$. We rewrite it in a form convenient to further analysis 
\begin{eqnarray}\label{nto0}
\left(1-\frac{\gamma\kappa g^{2}}{(1+\kappa^{2}g^{2})^{2}}\right)g''+\frac{1}{y}\left(1-\frac{\gamma\kappa g^{2}}{(1+\kappa^{2}g^{2})^{2}}\right)g'+\\
g \frac{\kappa\gamma(\kappa^{3}g^{2}-3)+2 \kappa^{2}(1+\kappa^{2}g^{2})^{2}}{(1+\kappa^{2} g^{2})^{3}}{g'}^{2} + \left(1-\frac{2 \kappa^{2}g^{2}}{1+\kappa^{2}g^{2}}\right)g = \sqrt{1+\kappa^{2}g^{2}}.\nonumber
\end{eqnarray}
The left-hand side of the above equation does not depend on the  specific potential and is peculiar to the Q-ball Ansatz in the baby Skyrme model. 
It is a differential equation of the second  order, hence for  given $y_{0}>0$, $g(y_{0})$ and $g'(y_{0})$ we should be able to set uniquely $g''(y_{0})$. But this is   not true for some values of $g$.
The above form of the equation makes this evident: the factors multiplying $g''$ and $g'$ may be simultaneously set to zero. We shall argue below that this lack of well-posedness gives rise to peakons. To this end we examine  solutions of~(\ref{nto0}) in a vicinity of $g_{r}$, where $g_{r}$ satisfies 
\begin{equation}\label{gam0}
1-\frac{\gamma\kappa g^{2}_{r}}{(1+\kappa^{2}g^{2}_{r})^{2}}=0.
\end{equation}
As for the number of roots of the above relation and their dependence on the parameters $\kappa$ and $\gamma$ see figure~\ref{roots1}. For convenience we introduce  a new function $\varepsilon(y)=g(y)-g_{r}$. In the vicinity of $\varepsilon=0$ the following equation holds
\begin{equation}\label{eq-approx}
\varepsilon \varepsilon''+\frac{1}{y}\varepsilon \varepsilon'+\alpha_{1}(\varepsilon')^{2}=\alpha_{2},
\end{equation}
where $\alpha$'s  are constants depending on $g_{r}$, $\gamma$ and $\kappa$. The equation is obtained from~(\ref{nto0}) by expansion around $g_{r}$ of the coefficients multiplying $g''$, $g'$,  ${g'}^{2}$ and the terms depending on $g$ only.  The leading terms are retained. The value of  $\alpha_{1}$ is entirely determined by the baby Skyrme model. It is equal to $1/2$ for $0<\kappa<\kappa_{max}$, for $\kappa_{max}$ it vanishes. In what follows we shall not put the actual value of $\alpha_{1}$ to make our considerations as general as possible. We shall see that the exact value of the parameter does not play an important role. The potential partially determines the value of $\alpha_{2}$. The equation~(\ref{eq-approx})  is valid unless one of its parameters ($\alpha_{1}$, $\alpha_{2}$) vanishes or explodes.  For a given $\gamma$ it happens for few discrete values of $\kappa$. Thus, the equation~(\ref{eq-approx}) is a generic one and we  concentrate on it.   \\
Let us state two immediate observations about the equation~(\ref{eq-approx}). First, the reflection $\varepsilon\to-\varepsilon$ does not alter the equation.  Second, rescaling of the variable $y\to \sqrt{|\alpha_{2}|}y$ makes  the constant term on the right-hand side in~(\ref{eq-approx}) equal to~$\pm1$. Hence, $\alpha_{2}$ sets the scale but its exact value is not crucial. The value of~$\alpha_{1}$ cannot be altered by rescaling of the function $\varepsilon$ or variable $y$.\\
Let us now analyze  the case  $\varepsilon'=0$. Then we have $ \varepsilon''=\alpha_{2}/\varepsilon$.   It means, $\varepsilon$ is repelled from the value $\varepsilon=0$  if $\alpha_{2}>0$ and  attracted to it if $\alpha_{2}<0$. To be more concrete, consider $\varepsilon<0$. If $\alpha_{2}>0$ then $\varepsilon$ has a maximum, otherwise minimum. Note, that the smaller absolute value at the extremum, the sharper the extremum is. This remark is valid even if $y=0$ is considered.\\
Let us note, that the equation~(\ref{eq-approx}) has a simple solution regular for $y=0$. It reads
\begin{displaymath}
\varepsilon(y)=\pm\sqrt{\frac{\alpha_{2}}{(1+\alpha_{1})}} y.
\end{displaymath}
The equation~(\ref{eq-approx}) is not analytically solvable in general. A great deal of information about the solutions may be inferred  from the equation. To this end the  substitution 
\begin{equation}
\varepsilon(y)=\left|w(y)\right|^{\frac{1}{1+\alpha_{1}}}
\end{equation}
is very useful. The function $w$ has only nonnegative values. The equation~(\ref{eq-approx}) turns into
\begin{equation}\label{Weq}
w''+\frac{1}{y}w'=-\frac{d}{dw}\left(-\alpha_{2}\frac{(1+\alpha_{1})^{2}}{2\alpha_{1}}w^{\frac{2\alpha_{1}}{\alpha_{1}+1}}\right).
\end{equation}
The above form of the  equation makes its mechanical interpretation natural. The equation~(\ref{Weq}) is  seen as a Newton equation for a particle in a potential subject to a friction (the term with the first derivative). Assume $2\alpha_{1}/(\alpha_{1}+1)>0$. Then, it is the sign of the coefficient $\alpha_{2}/\alpha_{1}$ that decides about the qualitative behavior of the solutions.  If it is negative, the particle is trapped in a potential well. Any solution eventually reaches the zero value then. This value is approached with a nonvanishing first derivative. Hence, the solution of~(\ref{eq-approx}) approaches zero as $\varepsilon\sim (y_{1}-y)^{1/(\alpha_{1}+1)}$. This gives rise to the exploding first derivative of $\varepsilon$ when $y\to y_{1}$. Thus, there is no  unambiguous way to continue the solution for $y>y_{1}$. 
Let us briefly interpret this finding in terms of the original equation~(\ref{nto0}). In a neighborhood of  $g_{r}$ any solution of~(\ref{nto0}) coincides with the corresponding solution of~(\ref{eq-approx}). Thus, any solution  coming close enough to $g_{r}$  has a dead-end since it  eventually reaches the value $g_{r}$ with infinite first derivative.\\
Now we discuss the case of  positive $\alpha_{2}/\alpha_{1}$ in equation~(\ref{Weq}). Then, the  potential has a global maximum for $w=0$. A fictitious particle can  climb  the  potential maximum if it is given enough (mechanical) energy. This gives rise to the exploding first derivative of $\varepsilon$ as described above. However,  this is not the only possible scenario. The particle may not have enough energy to reach the potential maximum. Then, it climbs for some time ($y$ is seen in this context as time), reaches its maximal accessible potential energy and finally dips. Of course, up to the initial data, the particle may just decrease its potential energy without approaching the maximum. A corresponding solution reads  asymptotically  $w\sim y^{\alpha_{1}+1}$, hence $\varepsilon(y)\sim y$.\\ 
There is also a solution separating the two just described families of solutions.  It is a solution  grazing the value $w=0$ at a point $y_{1}$. The function $w$ behaves then like $(y_{1}-y)^{\alpha_{1}+1}$, what gives $\varepsilon(y)\approx\pm\sqrt{\alpha_{2}/\alpha_{1}}|y_{1}-y|$. Precisely, the equation~(\ref{eq-approx}) is ambiguous for $\varepsilon=0$  and the behavior $\varepsilon\sim\pm(y-y_{1})$ is also correct. \\
Let us turn to the equation~(\ref{nto0}). Again, we concentrate on a strip around $g_{r}$, where the equation~(\ref{nto0}) turns effectively into~(\ref{eq-approx}). Consider  solutions entering this strip (thus the value of $g$ is set) at a given point. The further evolution depends on the derivative of $g$. If it is small, the function has an extremum and leaves the strip. The bigger absolute value of the derivative, the closer to $g_{r}$ the function $g$ approaches. For a critical value of the derivative the grazing solution appears. For larger values, the solutions touch $g_{r}$ with infinite first derivative. Figure~\ref{ilustr} illustrates the above discussion graphically.\\
In  light of the above analysis the appearance of peakons is just a question of amplitude. If a Q-ball profile function comes close to a root of the relation~(\ref{gam0}), then a peaklike solution is expected (if $\alpha_{2}/\alpha_{1}>0$). Let us stress once more that the above described structure does not emerge due to the V-shaped potential. It is  inherent in the Q-ball Ansatz in the baby Skyrme model. It would be interesting to see peakons in a broader perspective: what is their role in the whole model and what kind of nonlinearity makes them possible. \\
The peakons have been reported in the case of the spinning Q-balls, see~\cite{adam}. In this case the equation for the profile function is a bit more complicated. We have decided to work out the problem with $n=0$ as it seems  both generic and technically simple. What is more, some peakonlike nonspinning Q-balls are also present in the model. We shall see this in the   fifth section. In the next section some technical aspects peculiar to V-shaped potentials are described.\\
Finally, one more remark is in order. An analysis of equation~(\ref{nto0}) for large values of the function $g$ shows, that  solutions explode to infinity for finite $y$.  We are not about to delve into this here, though.
 
\section{The case $\kappa=0$ and power expansion}

Now, we are ready to look at Q-balls. We begin the analysis of the equation~(\ref{nto0}) with investigation of the case  $\kappa=0$. The equation has then the following form
\begin{equation}\label{sGeq}
g''+\frac{1}{y}g'+g=sign(g),
\end{equation}
coinciding with the equation for the Q-ball profile function in the signum-Gordon model. The relevant solutions in this model are well-known, see~\cite{ar} and~\cite{lis}. 
One can expect that the  Q-ball solutions of (\ref{nto0}) may be  expressed in the (traditional) power series of the  parameter $\kappa$: $g=g_{0}+\kappa g_{1}+\kappa^{2}g_{2}+\ldots$ for $\kappa$ ranging from zero to some maximal value. Note, that we can consider initial data with $\omega$ as large as needed. This means,  $\kappa$ may be arbitrarily close to zero. The boundary conditions supplementing both the equations~(\ref{nto0}) and~(\ref{sGeq}) are as follows. To ensure regularity of the solution at the origin it is necessary that $g'(0)=0$. The energy and charge are finite if $g(y)\to0$ if $y\to\infty$. The nonsmooth field force makes  the matching procedure indispensable, i.e. we need to find a point $y_{0}>0$, for which  $g(y_{0})=g'(y_{0})=0$. For $y<y_{0}$ the function $g$ is strictly positive, for $y>y_{0}$ the profile function vanishes. Let us now try to find a solution of the equation~(\ref{nto0}) in the following form
\begin{displaymath}
g=g_{0}+\kappa g_{1}+\kappa^{2}g_{2}+\ldots,
\end{displaymath}
where $g_{0}$ is a solution of~(\ref{sGeq}). To this end the above Ansatz is plugged into the equation. In the first order we obtain the relation
\begin{equation}\label{series1}
g_{1}''+\frac{1}{y}g'_{1}+g_{1}=-\gamma\left(\frac{g_{0}^{2}{g_{0}'}}{y}+g_{0}{g_{0}'}^{2}+g_{0}^{2}{g_{0}''}^{2}\right).
\end{equation}
It is solved using the Green function technique:
\begin{equation}\label{Green_sol}
g_{1}(y)=\int_{0}^{y}ds \ s \ G(y,s) F(g_{0}(s))+A_{1}J_{0}(y)+A_{2}Y_{0}(y),
\end{equation}
where $F$ denotes the r.h.s. in the equation~(\ref{series1}). $J_{0}$ and $Y_{0}$ are Bessel functions of the first and second kind solving  the homogeneous part of the equation~(\ref{series1}). Note, that $J_{0}(0)>0$ and $J'_{0}(0)=0$, but  $Y_{0}(y)\to\infty$ when $y\to 0$. Thus, due to the boundary conditions $A_{2}=0$ and  $A_{1}$ is a free parameter. We  use the following Green function
\begin{displaymath}
G(y,s)=\frac{J_{0}(s)Y_{0}(y)-Y_{0}(s)J_{0}(y)}{y\left(Y'_{0}(y)J_{0}(y)-J'_{0}(y)Y_{0}(y)\right)}.
\end{displaymath}
The integration in~(\ref{Green_sol}) gives a function behaving like $y^{2}\ln{y}$ at the origin. This Green function does not assume any condition at any other point $y>0$. This is why the freedom in adding a term proportional to $J_{0}$ is left (for further details consult~\cite{lis}). This freedom is necessary to make the function $g_{0}+\kappa g_{1}$ satisfy the boundary conditions.  One can  choose a different Green function with explicit dependence on the (unknown) matching point, see~\cite{ar}. However, it would not be helpful. To proceed, it has to be specified how to deal with the boundary conditions. We require that in any order of the power expansion the following conditions be satisfied: for any $m$ there be a point $y_{m}$ such, that 
\begin{displaymath}\begin{array}{c}
g_{0}(y_{m})+\kappa g_{1}(y_{m})+\ldots +\kappa^{m}g_{m}(y_{m})=0,\\
g'_{0}(y_{m})+\kappa g'_{1}(y_{m})+\ldots +\kappa^{m}g'_{m}(y_{m})=0.\\
\end{array}
\end{displaymath}
To this end the coefficient $A_{1}$ (and analogously in higher powers) has to vary with $\kappa$, making $g_{1}$ $\kappa$-dependent. Thus, the boundary conditions  make the expansion fail.  Another ambiguous decision to be made is how to understand $g_{0}$ for arguments larger than $y_{0}$. At this point the equation with nonsmooth force term is not uniquely determined, see~\cite{lis}. To sum up, the power series expansion is  just an abuse of notation.  Fortunately, we can resort to a more successful approximation.\\

\section{The Q-balls}
\begin{figure}
  \begin{center}
    \includegraphics[width=0.85\textwidth]{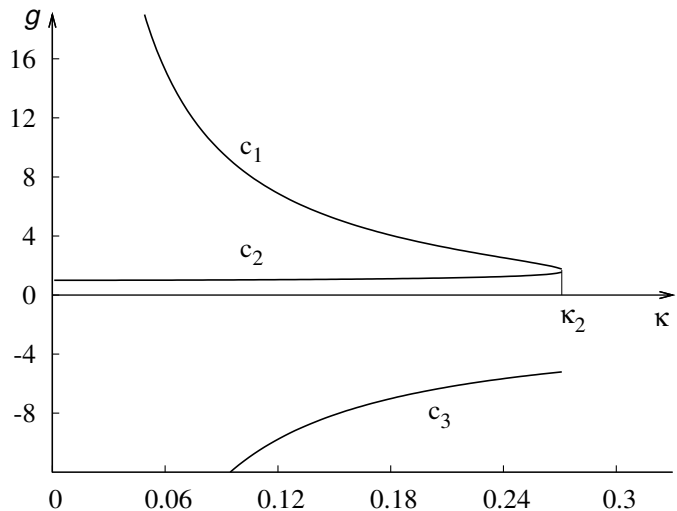}
    \caption{The three roots of the relation~(\ref{rooteq}).}
    \label{const-roots}
  \end{center}
 \end{figure}

\begin{figure}
  \begin{center}
  \begin{tabular}{cc}
    \includegraphics[width=0.49\textwidth]{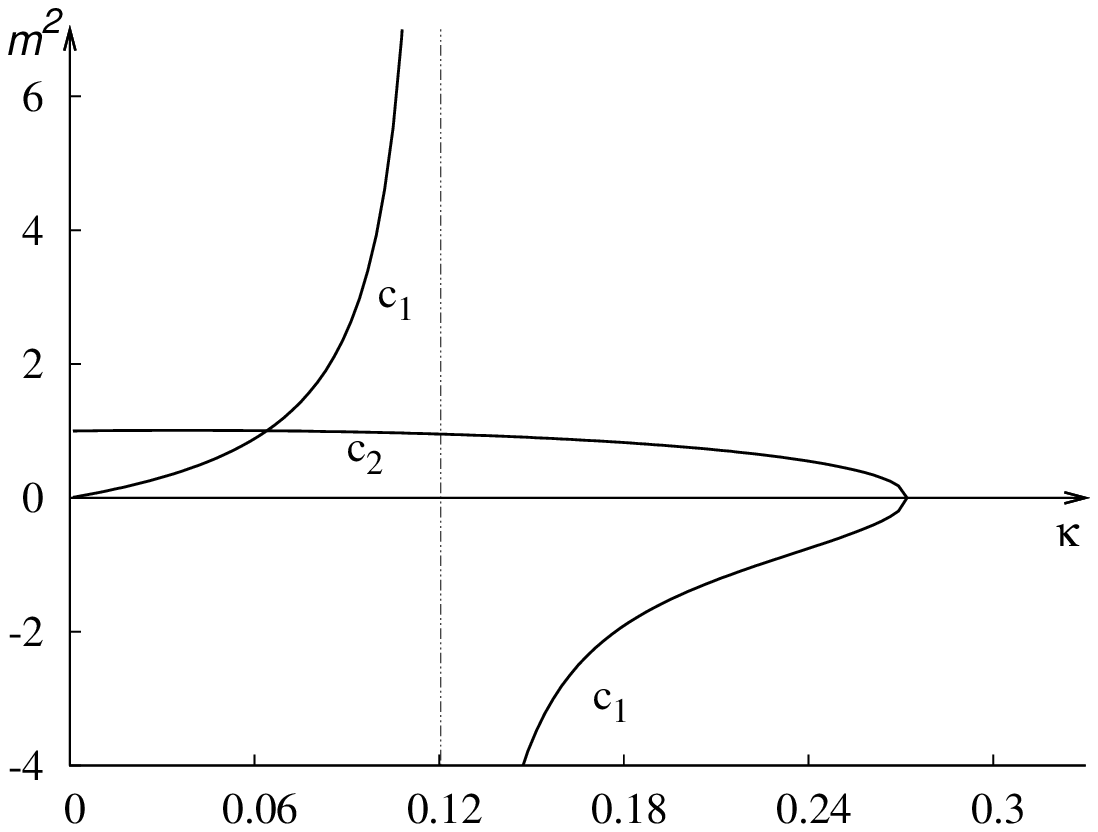}&
    \includegraphics[width=0.49\textwidth]{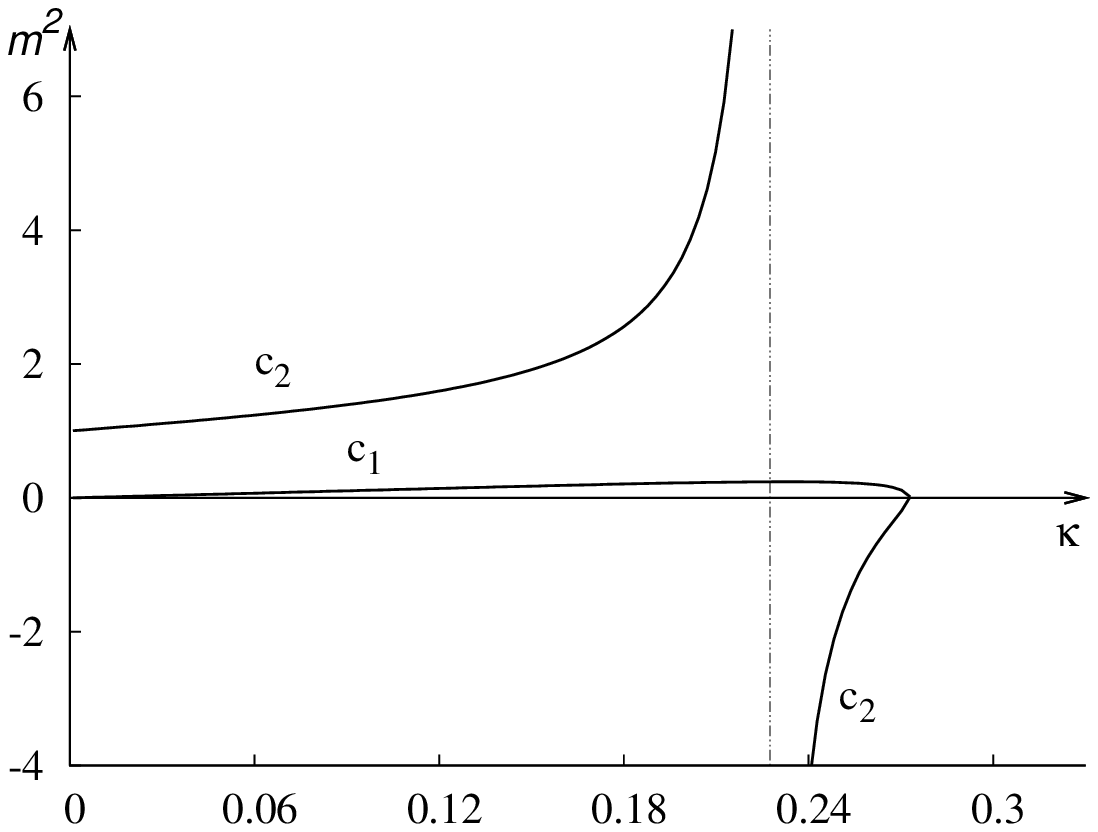}\\
    \end{tabular}
    \caption{Exemplary graphs showing the dependence of $m^{2}$ defined by formula~(\ref{mass}) on $\kappa$ for a given $\gamma$. $\gamma=0.5$ in the left panel, $\gamma=3.5$ in the right one. $m^{2}=0$ marks the merging point of the branches $c_{1}$ and $c_{2}$.}
     \label{masses11}
  \end{center}
 \end{figure}

\begin{figure}
  \begin{center}
   \begin{tabular}{cc}
    \includegraphics[width=0.49\textwidth]{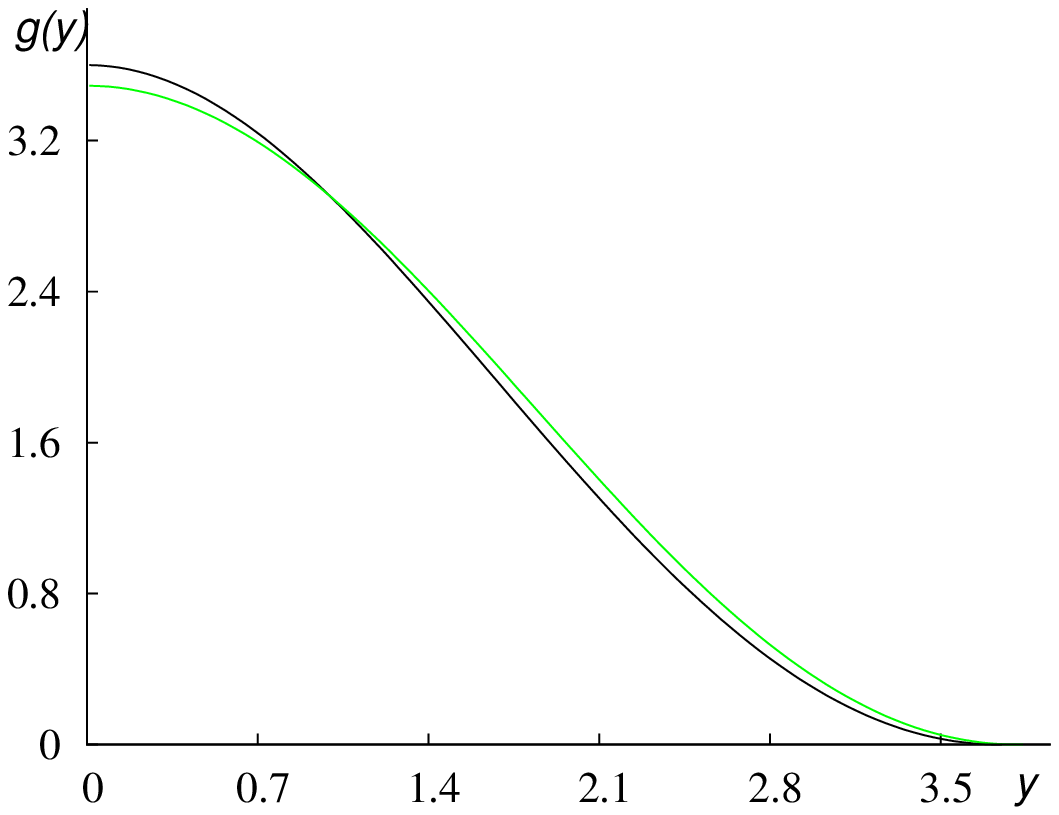} & \includegraphics[width=0.49\textwidth]{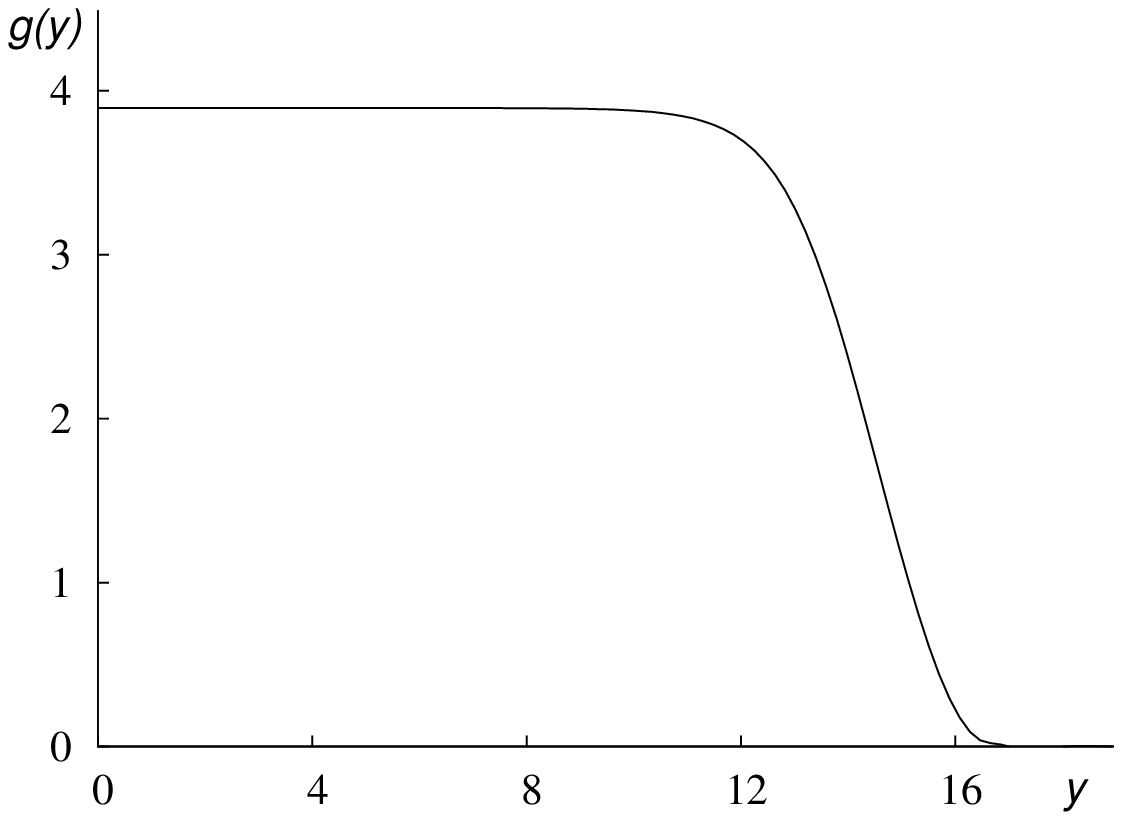}\\
    \includegraphics[width=0.49\textwidth]{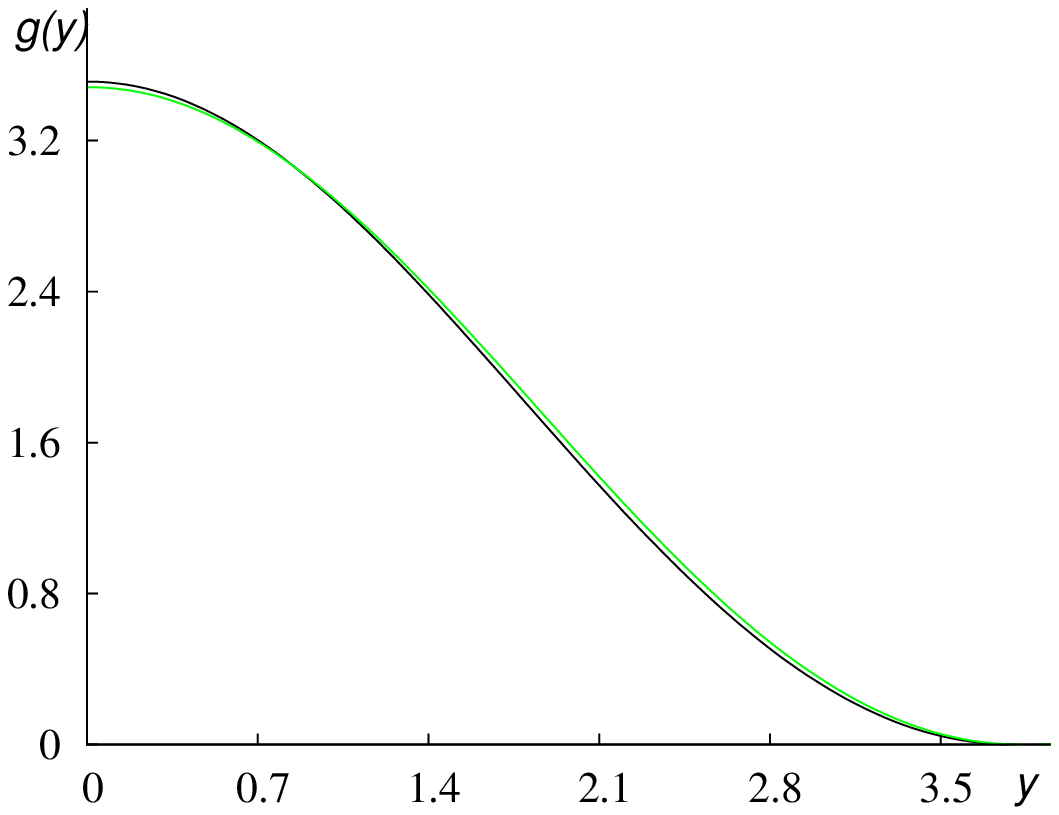} & \includegraphics[width=0.49\textwidth]{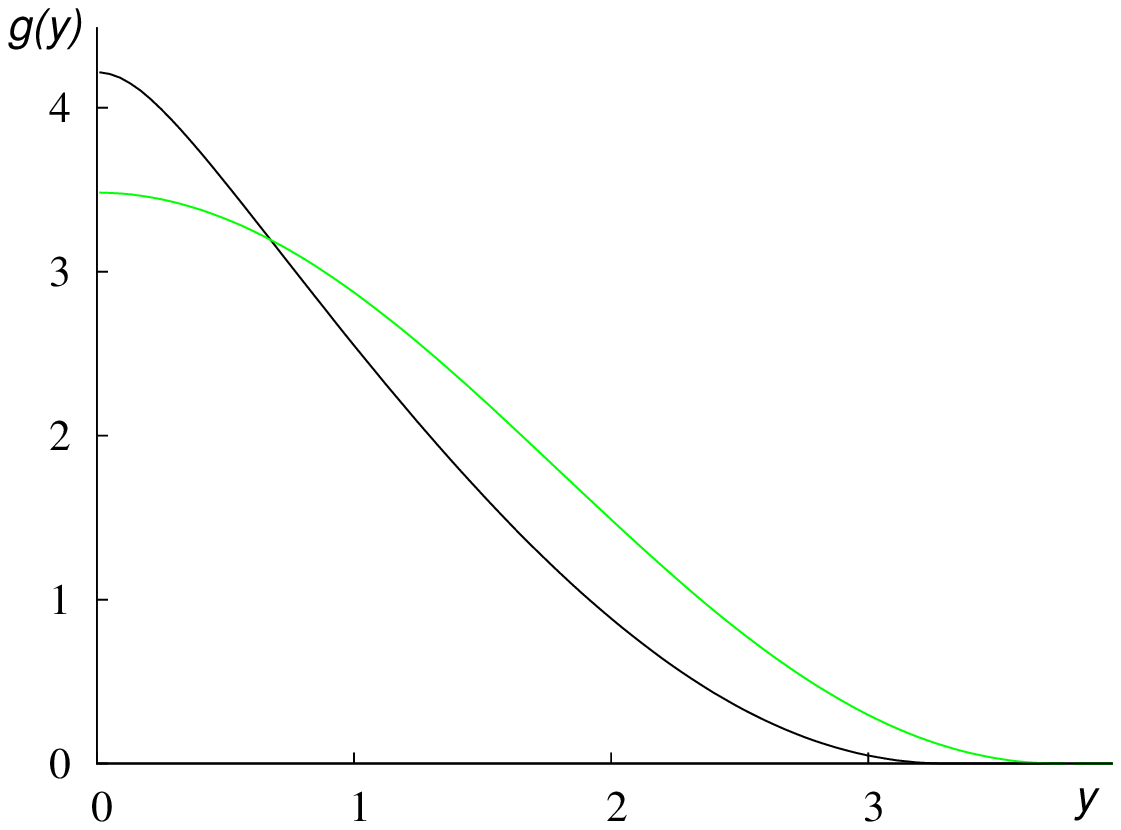}\\
    \end{tabular}
    \caption{The examples of the Q-ball profile functions. In the upper panel  $\gamma=0.5$, on the left picture $\kappa=0.03$, on the right $\kappa=0.185$. In the lower panel $\gamma=55$, $\kappa=0.0001$ on the left and $\kappa=0.00083$ on the right picture. The approximate solutions are drawn with green lines, the exact numerical solutions with the black ones.
    }\label{csol}
      \end{center}
 \end{figure}

In the case of an analytical potential giving rise to Q-balls there is an intuitive tool to cast a light on the relevant equation: the mechanical analogy. As described in the third section the equation for the profile function is then viewed as a Newton equation for a particle moving in an effective potential and subject to a friction, see e.g.~\cite{coleman}. A~\mbox{Q-ball} profile function is then usually a solution interpolating between three static solutions: one of them is stable,  two are not.
This picture is simplified in the signum-Gordon model. If $sign(g)$ in the equation~(\ref{sGeq}) is taken to be~$+1$, then the resulting equation has a simple structure: there is only one constant and stable solution. Any other solution may be viewed as its linear perturbation. It is the role of the boundary conditions to select a perturbation interpreted as a Q-ball solution. We shall adapt this idea to the baby Skyrme theory. Obviously, equation~(\ref{nto0}) is a nonlinear one and any linearization has a limited range of validity. \\
First let us find constant solutions of equation~(\ref{nto0}). They   satisfy the following relation
\begin{equation}\label{rooteq}
g=\frac{2 \kappa^{2}g^{3}}{1+\kappa^{2}g^{2}}+\sqrt{1+\kappa^{2}g^{2}}.
\end{equation}
It turns out, it  has zero, one, two or three real roots. Their structure illustrates  figure~\ref{const-roots}. There are three branches of solutions: $c_{1}$, $c_{2}$ and $c_{3}$. In what follows $c_{i}$ denotes a branch of roots in general or a specific root from the branch $c_{i}$. To avoid a possible ambiguity in the latter meaning we shall write $c_{i}(\kappa)$. For  $\kappa=0$ only one root exists and is equal to unity. In the limit $\kappa\to 0$ the other two roots explode to infinity as $\pm \kappa^{-1}$. 
The two  branches ($c_{1}$ and $c_{2}$) merge for $\kappa_{2}=\sqrt{2/27}$. The branch $c_3$ is negative for $\kappa\in(0,1)$ and for $\kappa>\kappa_{2}$ it is the only root of  the relation~(\ref{rooteq}). It is irrelevant in what follows and will not bother us any more. If $\kappa>1$, the equation has  no real roots. Note, that parameter $\gamma$ does not enter  into the relation~(\ref{rooteq}).\\
A linear perturbation around any of the constant solutions obeys the differential equation
\begin{equation}\label{FOperturb}
g_{1}''+\frac{1}{y}g'_{1}+m_{1}^{2}g_{1}=0,
\end{equation}
where 
\begin{equation}\label{mass}
m^{2}=\frac{4-(1+c^{2}\kappa^{2})\left( 3+c^{2}\kappa^{2}+c\kappa^{2}\sqrt{1+c^{2}\kappa^{2}}\right)}{1+c^{4}\kappa^{4}+c^{2}\kappa(2\kappa-\gamma)}.
\end{equation}
$c$ stands for a value of the perturbed constant solution. Roughly speaking, the solutions of~(\ref{FOperturb}) oscillate with decreasing amplitude if  $m^{2}>0$, otherwise they explode. In the  expression~(\ref{mass}) the parameter $\gamma$ plays an important role as   it makes the denominator vanish for some values of~$\kappa$. The infinite mass corresponds to a constant solution of the equation~(\ref{nto0}) being simultaneously a root of the relation~(\ref{gam0}). In this case the equation~(\ref{FOperturb}) is not the correct one.  To examine, how  solutions from its close neighborhood behave, an analysis of the equation~(\ref{eq-approx}) with $\alpha_{2}=0$ is required.\\
Figure~\ref{masses11} shows how  $m^{2}$ changes with $\kappa$. It makes clear, that for small $\gamma$ the infinite mass appears for the solutions from the branch $c_{1}$. The bigger $\gamma$, the smaller value of $c_{1}(\kappa)$  for which the linear approximation fails. Finally, for $\gamma>4\sqrt{6}/5$ the infinite mass appears on the lower branch $c_{2}$. When increasing $\gamma$, the denominator in~(\ref{mass}) vanishes for smaller and smaller $\kappa$.\\
Now we can describe the Q-balls present in the model. To this end we concentrate on a few generic profile functions.
Let us first consider the case of small~$\gamma$. To be more specific, we assume that $c_{2}(\kappa) \ll g_{r}(\kappa_{0})$. It means, the structure described in the third section does not influence the solutions.
We build the approximate Q-ball profile functions as a linearly perturbed $c_{2}(\kappa)$ solution. 
It is possible if $m^{2}>0$. In analogy to the signum-Gordon model the solution reads
\begin{equation}\label{app_sol}
g(y)=\left\{\begin{array}{lr}
c_{2}(\kappa)\left(1-\frac{J_{0}(m y)}{J_{0}(y_{0})}\right)&my<y_{0}\\
0&my>y_{0}\\
\end{array}\right. ,
\end{equation}
where the point $y_{0}\approx 3.8317$ is the smallest $y>0$ satisfying the condition $J_{0}'(y_{0})=0$ and $m=\sqrt{m^{2}}$. The fact, that the term $m^{2}$ weakly changes with $\kappa$, manifests itself in a scaling of the width of the solutions, since
\begin{displaymath}
y_{0}=m \omega r_{0}\approx \omega r_{0},
\end{displaymath}
where $r_{0}$ is a physical, not rescaled variable. Hence, $r_{0}\sim\omega^{-1}$. Such scaling was reported for spinning Q-balls in~\cite{adam}.\\
The approximation has an advantage:  it gives hints on the  regions in the parameter space, where no Q-ball profile function is possible. The approximation fails if the starting point $g(0)=c_{2}(1-J_{0}(0)/J_{0}(y_{0}))$ or ``ending point'' $g=0$ do not belong to the attraction region of the perturbed solution. Obviously it is the case, if $g(0)>c_{1}(\kappa)$.  This condition makes the range of allowed $\kappa$ values shrink to $(0, 0.186)$, where $\gamma$ is set to zero.  \\
For $\kappa$ close to zero the described approximation works well. The  constant solutions $c_{1}$ and $c_{3}$ are too far to anyhow modify the perturbation.  This is why the function~(\ref{app_sol}) is fairly close to the numerical solution, see figure~\ref{csol}.
The bigger $\kappa$ the worse the approximation works. The  explanation is rather easy: for bigger $\kappa$ the solutions start from a point in a close neighborhood of the constant solution $c_{1}$ with $m^{2}<0$. The profile functions stay then almost unchanged on a rather long interval, see figure~\ref{csol}.  Eventually, having left the vicinity of $c_{1}$ they oscillate around the solution $c_{2}$. Matching with the vacuum value is possible if $\kappa$ is  not too big. Taking too large value of $\kappa$ gives rise to a function oscillating around $c_{2}$ with an amplitude too small to reach the vacuum value.\\
Let us now move to larger values of the parameter $\gamma$. For small $\kappa$ both the constant solution $c_{1}(\kappa)$ and the value of $g_{r}$ are again irrelevant to a small perturbation around $c_{2}(\kappa)$. Thus, the approximation~(\ref{app_sol}) is fairly close to the actual solutions. For larger~$\kappa$ the Q-ball profile functions start at the origin with a value  close to a root of the relation~(\ref{gam0}). This results in sharpening  the maximum at the origin as described previously. Thus, we obtain  peakonlike solutions, see figure~\ref{csol}. It seems, that a maximal $\kappa$ admitting a Q-ball profile function emerges as a result of mutual approaching  $c_{1}$ and $c_{2}$ with increase of $\kappa$, analogously to the above described  case of large $\kappa$ and small $\gamma$. \\
We assume, that in our model  there is a region of small values of the parameter  $\kappa$ for each value of $\gamma$, where the  approximation~(\ref{app_sol}) works well. This is backed by the figures~1 and~3. They show, that for $\kappa$ small enough both the constant solutions $c_{1}$ and the roots of the equation~(\ref{gam0}) are far from $c_{2}$ and they are unlikely to influence a perturbation around this solution. This region shrinks when increasing $\gamma$ but  probably never  vanishes. It is noteworthy, that in case of a regular potential there is a maximal $\omega$ (minimal $\kappa$) allowing for the appearance of  Q-balls. This is why  we expect the solutions  not to exist for arbitrary large values of $\gamma$ then.\\ 
There has been some interest in the so-called  extreme baby Skyrme model, see e.g. \cite{speight2}, \cite{adam2}. It is a theory defined by a modification of the  Lagrangian~(\ref{lagrange}). The modification consists in omission of the first term in~(\ref{lagrange}). This theory is seen as a limit  $\beta\to\infty$ in the baby Skyrme model. For a particular potential it has been proven that there are no Q-balls in the extreme baby Skyrme model, see~\cite{adam2}. The just described mechanism of expulsion of the Q-balls from the theory with increase of  $\gamma$ suggests that there are no Q-balls in the extreme Skyrme model,  no matter what potential is taken into consideration.

\section{Conclusions}
In this paper we have considered in detail how the additional terms modifying the Laplace operator present in the baby Skyrme model influence the equation for the Q-ball profile function. These terms are responsible for appearance of peakons in the investigated theory. As we have mentioned above, it would be interesting to study the impact of this modification in a broader perspective. \\
Next we have demonstrated the failure of the power expansion in terms of a small parameter  in the theory with nonsmooth field force. The parameter dependence of the matching point may be blamed for this. Another way to characterize the parameter dependence has been devised. In this schema the Q-ball solutions appear as  relevant perturbations of a constant solution of the equation for the profile function.  Such a simplification  of a fairly complicated equation bears  not only qualitative understanding of the domain of existence of Q-balls, but also   a good quantitative description of the solutions in some range of parameters. \\
The work may be continued in many directions. Most obvious is a detailed investigation of the spinning Q-balls and gaining some qualitative (and quantitative if possible) understanding of these solutions. The role Q-balls may play in the baby  Skyrme model is also not known. To this end their stability should be studied. It would also be interesting to check if similar problems appear in the topologically nontrivial sectors.

\section{Acknowledgment}
I would like to thank  H. Arod\'z for fruitful discussions and urging me to write the paper. I  am also grateful to A. Wereszczy\'nski for pointing  out some issues  addressed in this article.


\begin{thebibliography}{11}
\bibitem{adam} C. Adam \emph{et al.}, Phys. Rev. \textbf{D80}, 105013 (2009).
\bibitem{speight1} J. Jaykka, M. Speight, Phys. Rev. \textbf{D82}, 125030 (2010).
\bibitem{zakrzewski} B. Piette, W. J. Zakrzewski, arXiv:hep-th/9710012v1;\\
 A. Kudryavtsev, B. Piette, W. J. Zakrzewski, Eur. Phys. J. \textbf{C1}, 333 (1998).
\bibitem{ar}  H. Arod\'z, J. Karkowski, Z. \'Swierczy\'nski, Phys. Rev. \textbf{D80}, 067702 (2009).
\bibitem{speight2} J. M. Speight,  J. Phys. \textbf{A43}, 405201 (2010).
\bibitem{rosenau} P. Rosenau, E. Kashdan, Phys. Rev. Lett. \textbf{104}, 034101 (2010).
\bibitem{ar3} H. Arod\'z, Z. \'Swierczy\'nski, arXiv:1106.3169v1.
\bibitem{lis} J. Lis, Acta Phys. Polon. \textbf{B41}, 629 (2010).
\bibitem{coleman} S. Coleman, Nucl. Phys. \textbf{B262}, 263 (1985).
\bibitem{adam2} C. Adam \emph{et al.}, Phys. Rev. \textbf{D81}, 085007 (2010).
\end{thebibliography}
\end{document}